\documentclass{article}
\usepackage{amssymb}
\usepackage{amsthm}
\usepackage[margin=1in]{geometry}
\usepackage[utf8]{inputenc}
\usepackage{mathtools}

\usepackage[singlelinecheck=false]{caption}
\usepackage{graphicx}
\usepackage{pgfplots}
\usepackage{pst-solides3d}
\usepackage{sidecap}
\usepackage{wrapfig}

\graphicspath{{./graphics/}}
\pgfplotsset{compat=1.18}
\sidecaptionvpos{figure}{m}

\usepackage{centernot}
\usepackage{extpfeil}
\usepackage{extarrows}
\usepackage{faktor}
\usepackage{mathrsfs}
\usepackage{xcolor}

\usepackage{hyperref}
\usepackage{cleveref}

\usepackage{tikz}
\usetikzlibrary{arrows}
\usetikzlibrary{automata}
\usetikzlibrary{calc}
\usetikzlibrary{cd}
\usetikzlibrary{decorations.markings}
\usetikzlibrary{decorations.pathmorphing}
\usetikzlibrary{shapes.geometric}

\tikzset{if/.code n args={3}{\pgfmathparse{#1}%
  \ifnum\pgfmathresult=1\pgfkeysalso{#2}\else\pgfkeysalso{#3}\fi}}
\tikzset{>=latex}
\tikzset{snake it/.style={decorate, decoration=snake}}
\tikzset{dbl/.style={>=stealth,
                     double,
                     double equal sign distance,
                     -implies,
                     shorten >=3pt,
                     shorten <=3pt}}          
\tikzstyle surjective=[postaction={decorate,decoration={markings, mark=at position .9 with {\arrow{latex}}}}]

\newcommand{\N}{\mathbb{N}}

\newcommand{\Z}{\mathbb{Z}}
\newcommand{\R}{\mathbb{R}}

\def\<{\langle}
\def\>{\rangle}

\newcommand{\lra}{\longrightarrow}

\DeclareMathOperator*{\argmin}{argmin}
\newcommand\Tr{\text{Tr}\ }

\newcommand{\sech}{\operatorname{sech}}

\DeclareMathOperator\arctanh{arctanh}

\makeatletter
\newcommand{\tpitchfork}{%
  \vbox{
    \baselineskip\z@skip
    \lineskip-.52ex
    \lineskiplimit\maxdimen
    \m@th
    \ialign{##\crcr\hidewidth\smash{$-$}\hidewidth\crcr$\pitchfork$\crcr}
  }%
}
\makeatother

\theoremstyle{definition}
\newtheorem{theorem}{Theorem}[section]

\newtheorem{definition}[theorem]{Definition}

\newtheorem{lemma}[theorem]{Lemma}
\newtheorem{proposition}[theorem]{Proposition}

\usepackage[minbibnames=2, backref=true]{biblatex} 
\usepackage{hyperref}
\usepackage{subcaption}

\title{\textbf{Correspondence Between Ising Machines\\ and Neural Networks}}
\author{Andrew G. Moore\\\normalsize Department of Mathematics, University of Texas at Austin\\
\normalsize\texttt{agmoore@utexas.edu}}
\date{October 31, 2025}

\begin{document}

\maketitle

\begin{abstract}
    Computation with the Ising model is central to future computing technologies like quantum annealing, adiabatic quantum computing, and thermodynamic classical computing. Traditionally, computed values have been equated with ground states. This paper generalizes computation with ground states to computation with spin averages, allowing computations to take place at high temperatures. It then introduces a systematic correspondence between Ising devices and neural networks and a simple method to run trained feed-forward neural networks on Ising-type hardware. Finally, a mathematical proof is offered that these implementations are always successful. 
\end{abstract}

\section{Introduction}

The Ising model holds a special place in the race to develop the computing technologies of the future. On one hand, it is the universal language of quantum annealing and adiabatic quantum computing, forcing a significant amount of quantum computing research to flow through the problem of Ising Hamiltonian design. On the other hand, emerging classical devices implementing the Ising model, such as coupled oscillators \cite{Cilasun2025}, memristors \cite{memristor} and, most recently, transistor-based p-bits \cite{jelinčič2025efficientprobabilistichardwarearchitecture}, promise a future of ultra low power compute-in-memory edge devices. No matter what hardware it is run on, the Ising model stands as the single most versatile physics-based method for solving NP-complete combinatorial optimization problems \cite{Lucas_2014, mohseni2022isingmachineshardwaresolvers}. However, the problem of designing Ising systems to execute computations is far from simple. Therefore, the single problem of developing new techniques to implement computation with the Ising model is of utmost importance to the development a wide variety of emerging and frontier technologies. 

Neural networks, on the other hand, are so omnipresent that their importance hardly requires elaboration. The connections between the Ising model and neural networks at a theoretical level have been noticed and studied since the origins of machine learning, but much still remains to be discovered about this relationship, and even more remains to be proven rigorously. Only a few papers have concerned themselves with running neural networks on Ising hardware, despite the fact that running a trained neural network on an Ising machine or quantum annealer would be very interesting practically as well as theoretically. This paper aims to patch that gap by providing not only a systematic technique for implementing certain feed-forward neural networks on Ising hardware, but also a rigorous mathematical proof of that method's correctness. 

\subsection{Main Results}

Traditionally, Ising computation operates at low temperatures, conceptually equating `solution' and `ground state' (\Cref{sec:intro}). This paper introduces a new approach to extracting solutions from Ising systems: instead of states, we observe the sign of spin averages (\Cref{sec:newidea}). This idea generalizes the notion of ground state solutions to all temperatures, allowing truly probabilistic noise-tolerant computation. 

The main result of this paper is that if one maps the parameters of a trained neural network with binary inputs and outputs to the coefficients of an Ising Hamiltonian, the resulting Ising system will behave like the neural network. A heuristic derivation of the high temperature case is provided in \Cref{sec:heuristic}). Practically, this result means that neural networks can be run natively on any hardware that implements the Ising model. Methods are provided, along with proofs of correctness, for implementing shallow tanh networks at high temperature (\Cref{thm:main}) and deep sign networks at low temperature (\Cref{thm:deep}).

The main technical idea is as follows. We observe that both $\tanh$-activated neural networks and the naive mean field approximation of the Ising model contain a hyperbolic tangent applied to an affine function (\Cref{sec:heuristic}). Though these two formulae are not exactly equal, we show that the error between them is at most little-o of a control parameter (\Cref{lem:full-accuracy}), while the magnitude of the circuit output is big-O of that same parameter. It follows that for the right parameter, the Ising system replicates the neural network (\Cref{sec:proof}). That control parameter is easily seen to be the temperature gradient, a common method for enforcing directionality in Ising circuitry.

Finally, we provide detailed computational experiments showing both the high and low temperature behavior for the case of 2-bit integer multiplication (\Cref{sec:shallow}). We also provide a quantized solution, with a heuristic discussion of quantization procedures (\Cref{sec:quantization}). 

\subsection{Other Work}

Traditional computing implements a complicated function (say, 16-bit integer multiplication) by introducing the abstraction layer of digital logic gates between the physical processes and the desired function. In contrast, a physics-based or physics-inspired computing device is a computing device that leverages physical dynamics and processes to perform computations directly \cite{PhysRevApplied.23.030001, aifer2025solvingcomputecrisisphysicsbased}. In practice, this definition encompasses a huge variety of novel approaches, but in our case it refers simply to the introduction of the Ising model, rather than logic gates, as the privileged abstraction layer. In other words, given a complicated Boolean function, we would like to know how to model it with Ising hardware. 

It is possible to design an Ising computer by implementing logic gates with Ising spins, then assembling them into digital logic, an approach we would call ``traditional logic via Ising.'' Many papers in the literature take this approach \cite{sequentialspinlogic, Caravelli_2020, Whitfield_2012, tsukiyama2024designingunitisingmodels}, as it is indisputably the shortest and clearest route to the implementation of a complicated function with Ising spins. This approach has proven effective for implementing functions such as multiplication and factoring, but necessarily shares many of the drawbacks of digital logic: while excellent for implementing well-understood mathematical functions, it is less than ideal for the problems tackled by artificial intelligence. A significant amount of research has also gone into designing Hamiltonians for natural Ising applications like combinatorial optimization problems \cite{Lucas_2014} and linear system solving \cite{novelrealnumberrepresentations}. However, much less attention has been given to the general design problem: how do we efficiently implement any Boolean function directly with Ising hardware in a natural way?

This paper will focus on this general design problem, in hopes that it will unlock the true potential of these frontier devices. We already have a great method of implementing arbitrary functions efficiently, even those that cannot be explicitly specified: the feed-forward neural network (FFNN). In fact, neural networks and their relatives are precisely the functions we would most like to offload to specialized inference hardware. It is therefore imperative that a way be found to run neural networks on Ising systems. 

Our correspondence, introduced in \Cref{sec:heuristic}, is based on mean field theory. Several papers in the past ten years have attempted to apply mean field theory to the analysis of feed-forward neural networks. In these cases the meaning is quite different: `mean field neural network' usually refers to a mean field approximation of neural network behavior as the number of neurons increases, not the idea that neural networks approximate the mean field behavior of another system. The closest to this work is  \cite{milletarí2019meanfieldtheoryactivation}, which modifies the statistical mechanics model directly, building in the directionality of information flow by analogy to renormalization, then shows that its mean-field behavior yields the $\tanh$-activated FFNN. We differ from this approach by focusing on the \textit{approximate} relationship between the two \textit{unmodified} models, and especially the use of the feed-forward model as a design surrogate for the loopy model. To the knowledge of the author, this approach has not yet been analyzed. One paper \cite{stosic2022isingmodelsdeepneural} does consider an Ising model directly corresponding to a trained feed-forward neural network (as well as a transformer), but not in the context of attempting to make the Ising system behave like, implement, or simulate the neural network. Instead, in that work the thermodynamics of the Ising model are used as an analytical tool to examine the internal structure of the network weights. 

Another closely related paper is \cite{traininganisingmachinewithequilibriumpropagation}, which also considers a direct correspondence between neural networks and Ising circuits. That work, however, differs from ours in every other respect: instead of using a trained neural network to program the weights of an Ising circuit, it recommends training the neural network directly in the context of Ising hardware. This produces good practical results, but unlike our method it is not amenable to theoretical performance guarantees, due to the difficulty involved in analyzing the convergence of message passing algorithms on cyclic graphs \cite{Wainwright2008GraphicalME}. \cite{song2023trainingmultilayerneuralnetworks} implements a similar idea, also training quantized Ising neural networks directly, using a secondary Ising machine for training. Recently, diffusion-denoising algorithms have been successfully implemented on classical hardware that emulates the Potts model \cite{jelinčič2025efficientprobabilistichardwarearchitecture}, indicating a growing interest in mapping artificial intelligence models directly to quadratic energy-based hardware. In the present work, we will focus on the theoretical properties of the correspondence, allowing us to provide a rigorous proof of correctness. Rather than analyzing a complex training algorithm, we only have to show that our mapping preserves the correctness of an already trained network.

\section{The Ising Model as a Computing Device}

We will begin by introducing the definition of an Ising circuit, as well as reviewing the traditional approach to Ising computation, namely the prescription of ground states.
Following that, we propose a new approach: considering Ising circuits as observed through rounded time-averages at high temperatures rather than by sampling of single states at low temperatures. The remainder of the paper will focus mainly on the development of design methods for Ising circuits of this new type. 

\subsection{What is an Ising Circuit?}

We consider a finite set of spins $G$, each of which has state space $\Sigma = \{\pm 1\}$. Any state of the system is therefore an element of $\Sigma^{|G|}$. The system is equipped with a quadratic Hamiltonian \begin{align}
    H_{h,J}(\sigma)=-\langle h, \sigma\rangle - \frac{1}{2}\langle \sigma, J\sigma\rangle,
\end{align}
where $\sigma \in\Sigma^{|G|}$ is any configuration, $h$ is the vector of local magnetic field strengths, and $J$ the symmetric zero-diagonal interaction matrix. The system carries an equilibrium probability distribution on its configuration spaced called the \textit{Boltzmann distribution}, denoted
\begin{align}
    \mathbb{P}[\sigma] = \frac{1}{Z}\exp(-\beta H_{h,J}(\sigma)),
\end{align}
where $\beta = 1/T$ is the inverse temperature (in natural units) and $Z = \Tr \exp(-\beta H_{h,J})$
is a normalizing factor called the \textit{partition function}. In the high temperature limit $(\beta \to 0)$ the Boltzmann distribution becomes uniform, while in the low temperature limit $\beta \to\infty$ it concentrates at the lowest energy state, or \textit{ground state}. 

So far, what we have described is merely the standard Ising model. In order to identify this system as a circuit, we choose disjoint subsets $X, Y \subset G$. $X$ is called the set of \textit{input spins} and $Y$ is called the set of \textit{output spins}. The remaining spins $A := G \setminus \{X \cup Y\}$ are called the \textit{auxiliary spins}. It will be convenient to denote $H(\sigma)$ as $H(x,y,a)$, where $x \in \Sigma^X$, $y \in \Sigma^Y$, and $a \in \Sigma^A$. We will assume that we may fix the set of input spins. In other words, $X$ can be held constant at any pattern $x \in \Sigma^X$. To operate the circuit, we hold $X$ fixed at a desired input $x$, and somehow observe $Y$ to extract our result (more will be said on this in a moment). Therefore, when working with Ising circuits we are usually interested in the family of $2^{|X|}$ conditional Boltzmann distributions $\mathbb{P}[(y,a) | X=x]$ for each $x$, rather than the full Boltzmann distribution, since $X$ is controlled and does not evolve freely. Similarly, we are concerned with the conditional expectations $\langle (y,a)|X=x\rangle$, not the full expectations $\langle (x,y,a)\rangle$. In lengthy proofs, we will sometimes use $\langle \cdot \rangle$ to mean the expectation conditioned on $X=x$, when it is clear that we are working with a single fixed input.

\subsection{Traditional Approach: Ground States as Solutions}
\label{sec:intro}

The `traditional' approach to implementing computation with Ising systems is to draw an equivalence between ground states and solutions. If the ground state is the solution, the solution can be obtained by cooling the system. In the context of Ising circuits, the approach 
is to use a low temperature and simply observe the state of $Y$, instantaneously and synchronously, to find the output. In other words, the correct output should be the conditional ground state given the input. This is equivalent to a (near) zero-temperature conditional Boltzmann machine. For the moment, we are assuming that our device is at thermal equilibrium, and will not concern ourselves with dynamical issues like autocorrelation time and  annealing rate.

The problem of designing the Ising circuit can then be thought of as `programming' the input-output pairs defining a desired function as the ground states of a quadratic Hamiltonian, conditioned on the input spins. Suppose $f:\Sigma^X \lra \Sigma^Y$ is some Boolean function we would like to implement as a circuit. We say that $H$ \textit{implements $f$ in the zero-temperature sense} if $Y = f(X)$ a.s. in the low temperature limit. Stated more precisely, this is the condition
\begin{align}
    \lim_{\beta \rightarrow \infty} \mathbb{P}[Y=f(x) | X = x] = 1 \forall x \in \Sigma^X,
\end{align}
where states are distributed according to the Boltzmann distribution on $H$ and $\beta$. Up to infinitesimal perturbation of the coefficients of $H$, this condition is equivalent to stating that there exists a function $g : \Sigma^X \lra \Sigma^A$ such that
\begin{align}
\label{eq:argmin}
    (x, f(x), g(x)) = \argmin_{y \in \Sigma^Y, a \in \Sigma^A} H(x,y,a) \forall x \in \Sigma^X.
\end{align}
The value of the argmin is precisely the ground state of the system conditioned on $X=x$. 
\Cref{eq:argmin} is equivalent to a system of linear inequalities. Therefore, if $g$ is known, the coefficients of $H$ can be determined by linear programming (LP). The problem is then broken into two stages: first, find a $g$ which renders the LP problem feasible; second, solve the LP problem. It is also possible to find $H$ by using the training algorithms usually used for conditional restricted Boltzmann machines \cite{mnih2012conditionalrestrictedboltzmannmachines}. This has the advantage of avoiding the difficulty of finding a feasible choice of $g$, but in practice it is even slower and more unreliable than the mixed binary integer-linear programming problem described above, at least once algorithms for the latter are optimized. This is because the contrastive divergence algorithm easily gets stuck and struggles to get every input-output pair correct. The weakness of the Boltzmann machine formalism for modeling Ising circuitry was also noted by \cite{traininganisingmachinewithequilibriumpropagation}. 

We have previously dedicated substantial work \cite{moore2025geometrictheoryisingmachines, martin2023designgeneralpurposeminimalauxiliary} to investigating the low-temperature Ising design problem just described. However, framing the problem in this way has some drawbacks. The first issue is purely computational: even once a feasible $g$ is found, determining the coefficients of $H$ requires solving a linear programming problem whose size is $\mathcal{O}(k^2 2^k)$, where $k$ is the total number of spins. While we have found search algorithms and heuristics for finding viable choices of $g$, this procedure is completely impractical for circuits larger than $4$-bit integer multiplication.
By contrast, neural network architectures can learn precise algorithms applicable to the entire input space from stochastic sampling of training inputs \cite{mccracken2025uncoveringuniversalabstractalgorithm, musat2024clusteringalignmentunderstandingtraining}, which opens the possibility of circumventing this curse of dimensionality. 

\subsection{The Alternative: Rounded Average States as Solutions}
\label{sec:newidea}

Must `ground state' be a synonym of `solution'? At low temperatures, the system concentrates at its ground state, so this notion makes sense. There is also a significant amount of literature regarding encoding the solution to combinatorial optimization problems in Ising ground states. However, this does not mean \textit{a priori} that the two ideas must be equivalent. If ground states are solutions and we extract information from the system by observing states, then when the temperature is above zero at time of observation we experience non-zero failure probability. Therefore, in this setup, randomness at observation time is directly associated with failure. Annealing circuits indeed require noise to search for the minimum, but also require its suppression in order to observe that minimum reliably. Why, given a probabilistic system, should our method of solution center around making it more deterministic? What if we could accept a high temperature at all times, and yet still extract useful information? Could a circuit exist which performed better when the temperature at observation time was higher?

At high temperatures, we can still reliably observe spin averages. Suppose that our hardware equilibrates rapidly relative to the observer, allowing us to observe time-averaged spin states. Choosing to observe averages allows the output of our circuits to be, in principle, real numbers in $[-1, 1]$. However, since we are modeling a Boolean function, we will clamp averages to the nearest spin to obtain a binary result. This renders the circuit directly comparable to the zero-temperature circuit, and creates valuable wiggle room in the design process, since the exact value of the spin average will not matter. Now, we can rephrase the design problem as follows: find $H$ and $\beta$ such that for all inputs $x \in \Sigma^N$, $\text{sign}(\langle Y | X=x\rangle) = f(x)$
where sign is the per-spin sign function. Any positive average is considered a $1$ and any negative average is considered a $-1$. The magnitude of $\langle Y | X=x\rangle$ indicates confidence, and can be used to determine how long we must observe before being sure that the sign is correct to within a specified error rate. Since this definition of `solution' will be used constantly for the rest of this paper, we will state it again:

\begin{definition}[High-Temperature Solution]
    Consider an Ising system with Hamiltonian $H$ and inverse temperature $\beta$. Let $f$ be a Boolean function $\Sigma^{|X|} \lra \Sigma^{|Y|}$. We say that $(H,\beta)$ \textit{implements $f$ in the high-temperature sense}, or simply \textit{solves $f$}, if 
    \begin{align}
        \text{sign}(\langle Y | X=x\rangle_{H,\beta}) = f(x) && \forall x \in \Sigma^{|X|},
    \end{align}
    where $\langle \cdot \rangle$ indicates expectation taken over the Boltzmann distribution. 
\end{definition}

The new definition clearly generalizes the previous definition to positive temperatures, though there is an important subtlety. A zero-temperature \textit{solution} $H$ is a high-temperature \textit{solution} (for low enough finite temperature). Therefore, zero-temperature \textit{solvability} implies high-temperature \textit{solvability}. The converse to the first statement is not true: a high temperature \textit{solution} $H$ is not always a zero-temperature solution. However, we do not know whether the converse to the second statement is true, that is, whether there exist functions solvable at high temperature which are not solvable at zero temperature, for a fixed number of auxiliary spins. In general, proving that any circuit is impossible with a fixed nonzero number of auxiliary spins, in either the high or zero temperature sense, is quite difficult, even when it is intuitively obvious. This is due to the exponentially large number of potential auxiliary states and the lack of mathematical tractability in relating traditional notions of circuit complexity to spin systems involving loops. 

\begin{proposition}[Relationship of High-Temperature and Zero-Temperature Solutions]
\
    \begin{enumerate}
        \item Fix a set of spins, $X$, $Y$, and function $f : \Sigma^{|X|} \lra \Sigma^{|Y|}$. Then, if $H$ implements $f$ in the zero-temperature sense, there exists $\beta^* > 0$ such that for all $\beta > \beta^*$, $(H,\beta)$ implements $f$ in the high-temperature sense.
        \item The converse is not true: there exists a set of spins, $X$, $Y$, $f$, $H$, and $\beta$ such that $(H,\beta)$ implements $f$ in the high-temperature sense, but $H$ does not implement $f$ in the zero-temperature sense. 
    \end{enumerate}
    \begin{proof}
        Part (a) follows immediately from Laplace's principle. Part (b) is proven by example in \Cref{sec:shallow}.
    \end{proof}
\end{proposition}

This approach is more plausible for some hardware implementations than others. Quantum annealing is perhaps the most obviously attractive, because it obeys the Boltzmann distribution at a chosen temperature determined by the strength of the transverse field. Furthermore, the high-temperature perspective allows us in principle to anneal more quickly and stop the annealing earlier, potentially speeding up computation. Classical systems which switch quickly such as magnetic tunnel junctions \cite{patel2024aiguidedcodesignframeworknovel}, combined with averaging circuits for observation, are also promising. Systems which rely on low temperatures, such as adiabatic quantum computing, and those which obey the Ising model exactly only at low temperatures, such as coupled oscillators and other Kuramoto-type systems, would require extra work to apply this conceptual framework. We will assume that we are working with an idealized Ising system whose states are distributed precisely according to the Boltzmann distribution. How exactly this is achieved, and with what degree of precision, is firmly in the realm of hardware and out of the scope of this work. For most of this paper, we will focus exclusively on high-temperature solutions, returning to the zero-temperature case only in \Cref{sec:deep}. 

The question now becomes how we `train' the high-temperature setup. In other words, how do we decide the weights for the spin interactions so that at some specified $\beta$, the average values of the output spins have the correct signs for each input? Facially the model resembles a conditional Boltzmann machine \cite{mnih2012conditionalrestrictedboltzmannmachines}, but these models optimize for the observation of states, not the observation of averages. Determining the interaction strengths from observed averages appears commonly in the context of variational inference on Markov Random Fields \cite{Wainwright2008GraphicalME}, Hamiltonian learning \cite{Ott_2024}, and inverse Ising statistics \cite{Nguyen_2017}. The known methods for these problems focus on approximately reconstructing the entropy or partition function in a computationally tractable way, then computing the approximate mapping from averages to parameters. One such method, equilibrium propagation, has actually been implemented for training Ising-based neural networks \cite{traininganisingmachinewithequilibriumpropagation}. 

Note, however, that in our setting we do not care about the actual value of the spin averages, as long as they have the right sign and are not too close to zero. We also do not specify the correlation averages, nor anything about the hidden spins. In some sense, these statistical methods are overkill when the goal is modeling a binary function. One common system, however, presents a compelling parallel to our setup, while remaining very straightforward to train: the standard shallow hyperbolic tangent neural network. While superficially quite unlike the Ising model, it turns out that this classic system is quite related to our new model of circuits.

\section{Correspondence Between Ising Circuits and Neural Networks}
\label{sec:heuristic}

This section will provide a non-rigorous version of the derivation of \Cref{thm:main}. Rigorous proofs will be provided in \Cref{sec:main}.

\vspace{0.15cm}
\noindent
Consider the graph topology of a feed-forward neural network with $L+1$ layers and $M_\ell$ spins in layer $\ell$. We will denote the $i$th spin in the $\ell$th layer by $\sigma^{(\ell)}_i$ and the connection weight between $\sigma^{(\ell-1)}_i$ and $\sigma^{(\ell)}_j$ as $J^{(\ell)}_{ij}$. All other connections have coupling strength zero, enforcing the layered structure. We can also include local magnetic fields $h^{(\ell)}_i$. We will refer to $\sigma^{(0)}$ as the inputs $X$ and $\sigma^{(L)}$ as the outputs $Y$, with the intermediate layers being referred to as hidden or auxiliary. Elementary Ising theory tells us that, assuming all spins except $\sigma^{(\ell)}_i$ are fixed, 
\begin{align}
    \langle \sigma^{(\ell)}_i\rangle = \tanh\left(\beta\sum_{j=1}^{M_{\ell-1}} J^{(\ell)}_{ij} \sigma^{(\ell-1)}_j + \beta\sum_{j=1}^{M_{\ell+1}} J^{(\ell+1)}_{ji} \sigma^{(\ell+1)}_j + \beta h^{(\ell)}_i\right).
\end{align}
In other words, the average state of a spin is the hyperbolic tangent of an affine function of the other spins, when those spins are fixed. This should remind us of a neural network with tanh activation. However, the other spins being fixed spoils the analogy somewhat. The analogy would work better if the spin averages were related \textit{to each other} by the tanh of an affine function. The naive mean field (NMF) theory of the Ising system gives us exactly that sort of (approximate) statement: 
\begin{align}
    \langle \sigma^{(\ell)}\rangle \simeq \tanh\left(\beta J^{(\ell)}\langle \sigma^{(\ell-1)} \rangle + \beta (J^{(\ell+1)})^T \langle \sigma^{(\ell + 1)}\rangle + \beta h^{(\ell)}\right) .
\end{align}
If, furthermore, the $J^{(\ell+1)}$ are very small compared to the $J^{(\ell)}$, then 
\begin{align}
\label{eq:ansatz-directional}
    \langle \sigma^{(\ell)}\rangle 
     \simeq \tanh\left(\beta J^{(\ell)}\langle \sigma^{(\ell-1)} \rangle + \beta h^{(\ell)}\right) .
\end{align}
\Cref{eq:ansatz-directional} is states that the spin averages are roughly given by the forward propagation formula for a neural network using $\tanh$ activation with weights $W^{(\ell)} = \beta J^{(\ell)}$ and biases $b = \beta h$, where the value of each neuron is the average value of the spin. 

We have found a way to make an Ising system look like a neural network. Now, suppose that we wanted to go the other direction: if we had neural network weights $W,b$, how would we find the Ising system  $\beta, J, h$ corresponding to it? Assume that the $W^{(\ell)}$ are of roughly similar magnitude. If we set $\beta = 1, J^{(\ell)} = W^{(\ell)}$, we will obtain an Ising system, but its behavior will not approximate that of the neural network because the approximation \Cref{eq:ansatz-directional} is not valid. To make the connection, we need to force the later layers to have lower magnitude. In other words, we need to introduce a \textit{temperature gradient} in which the input is the `cold' end and the output is the `hot' end. This terminology is implied by the fact that in the Ising model, larger interaction magnitudes are equivalent to higher $\beta$, that is to say to lower temperatures. Decreasing the interaction magnitudes of later layers therefore corresponds to increasing their temperature. 

The natural way to do this is to introduce a parameter $0 < \delta < 0$ and set $(J^{(\ell)}, h^{(\ell)}) = \delta^{\ell-1} (W^{(\ell)}, b^{(\ell)})$. This scales $J^{(\ell+1)}$ by a factor of $\delta$ relative to $J^{(\ell)}$ for all $\ell$, and if $\delta$ is small enough, then $J^{(\ell+1)}$ are very small compared to the $J^{(\ell)}$, so \Cref{eq:ansatz-directional} is valid. This technique is very common in the literature \cite{sequentialspinlogic, Caravelli_2020}. This leads us to formalize our correspondence between Ising systems and neural networks:

\begin{definition}[Correspondence Principle]
\label{def:cor}
Given a choice of two free parameters, an inverse temperature $\beta >0$ and a temperature gradient $1\geq\delta >0$, 
a feed-forward neural network with $L+1$ layers and parameters $\{W^{(\ell)},b^{(\ell)}\}_{1\leq\ell\leq L}$ corresponds to an Ising system on the same underlying graph topology equipped with inverse temperature $\beta$ and Hamiltonian
\begin{align}
    H_{\delta}(\sigma) = -\sum_{\ell=1}^{L} \delta^{\ell-1}\left[\langle b^{(\ell)},\sigma^{(\ell)}\rangle + \langle \sigma^{(\ell)}, W^{(\ell)}\sigma^{(\ell-1)}\rangle\right].
\end{align}
Since the multipartite graph structure is known, it is usually more convenient to speak of the Ising system as having couplings $J^{(\ell)} = \delta^{\ell-1}W^{(\ell)}$ and local field strengths $h^{(\ell)} = \delta^{\ell-1}b^{(\ell)}$. 
\end{definition}

However, we have introduced a problem. Intuitively, the issue is that the network was trained with an implied inverse temperature $\beta=1$ at every layer, but with a temperature gradient that assumption must be violated at every layer but one. If $L$ is large, then we will be forced to distort the activation functions away from $\tanh$. The correspondence effectively deforms the activation function via the mapping $\tanh(A) \to \tanh(\delta^{\ell-1}(\beta A+\mathcal{O}(\delta)))$, where $A$ is the incoming pre-activation. 
This distortion of $A$ by a scaling factor of $\delta^{\ell-1}\beta$ does not change the sign, but it does change the value in a nonlinear way. The distortion is thus acceptable on the last layer, but if it occurs at any previous layers, the correspondence will become inexact and intractable: no matter the choice of $\beta$, some layers will have steeper activations and others shallower relative to the neural network model, throwing everything into confusion.

However, when $L=2$, that is the case of a shallow neural network, our correspondence survives: if we set $\beta = 1$ and pick a small $\delta$ the behavior of the hidden layer exactly mimics the neural network, while the sign of the output layer remains unchanged. The shallow neural network function may be written with the convenient notation
\begin{align}
    \phi_{W,b}(s) := \tanh\left( W^{(2)} \tanh\left(W^{(1)} s + h^{(1)}\right) + h^{(2)}\right).
\end{align}
Expressed in mean field equations, the network is mapped to an Ising system roughly obeying
\begin{align}
    \langle \sigma^{(0)}\rangle &= \sigma^{(0)} \text{ fixed},
    \\
    \langle \sigma^{(1)}\rangle &\simeq \tanh\left(W^{(1)} \sigma^{(0)} + \delta (W^{(2)})^T \langle \sigma^{(2)}\rangle + h^{(1)}\right),
    \\
    \langle \sigma^{(2)}\rangle &\simeq \tanh\left(\delta( W^{(2)}\langle \sigma^{(1)} \rangle + h^{(2)})\right) .
\end{align}
As $\delta \rightarrow 0$, we have the asymptotic result
\begin{align}
    \langle \sigma^{(1)}\rangle &\sim \tanh\left(W^{(1)} \sigma^{(0)} + h^{(1)}\right), 
    \\
    \langle \sigma^{(2)}\rangle &\sim \delta\left( W^{(2)}\langle \sigma^{(1)} \rangle + h^{(2)}\right) .
\end{align}
Therefore, if $\delta$ is small enough to treat these asymptotic relationships as equalities, we replicate the signs of the outputs of the neural network:
\begin{align}
    \text{sgn}(\langle \sigma^{(2)}\rangle) 
    = \text{sgn}\left(\delta\left( W^{(2)}\langle \sigma^{(1)} \rangle + h^{(2)}\right)\right) 
    = \text{sgn}\left(\tanh\left( W^{(2)}\langle \sigma^{(1)} \rangle + h^{(2)}\right)\right)
    \\
    = \text{sgn}\left(\tanh\left( W^{(2)} \tanh\left(W^{(1)} \sigma^{(0)} + h^{(1)}\right) + h^{(2)}\right)\right) 
    = \text{sgn}(\phi_{W,b}(\sigma^{(0)})).
\end{align}
Therefore if the neural network was trained to approximate the function $f$, we obtain an Ising circuit which implements $f$ in the high-temperature sense. This line of argument strongly suggests that we can design Ising circuits by training a shallow $\tanh$ neural network to predict the function, copying the weights to the couplings of the Ising model, and introducing a strong enough temperature gradient.

The correspondence is based on the trivial observation that both tanh-activated neural networks and naive mean field theory contain a hyperbolic tangent applied to an affine function. The reader might therefore be inclined to infer that the hyperbolic tangent was originally introduced as a neural network activation function because of its role in mean-field spin systems. However, this was probably not the reason for its use. Early neural networks used $\tanh$ as an alternative to the sigmoid function, itself an alternative to the sign function. Historically, hyperbolic tangent was introduced due to its convenient range, high degree of regularity, simple derivatives, and computational tractability. As early as the 1980s, Amit, Gutfreund, Sompolinsky, and Hopfield were leveraging the statistical mechanics connection to analyze recurrent spin systems understood as neural networks. However, this work came long after the introduction of $\tanh$ as an activation function was not applied to feed-forward networks, only to recurrent systems whose relationship to spin models is more exact. 

Of course, we made some hasty assumptions. Firstly, while it is common knowledge that mean field theory is accurate in the high temperature limit, is is really true that we can use mean field theory in this case, even though the temperature of the first layer is fixed? Secondly, note that the output spin averages are becoming small in $\delta$. How do we know that the approximation error from the mean field theory or the directional flow ansatz doesn't dominate the magnitude of the output, resulting in sign errors at small $\delta$? In fact, these worries are unfounded and this design method always works (that is \Cref{thm:main}). Before addressing this precisely, however, we will investigate a particular example of a circuit which is large enough to exhibit interesting and novel behavior, but small enough that the spin averages can be calculated directly. 

\section{Simulated Experiments}
\label{sec:shallow}

This section concerns experimental implementation of a toy model, applying our correspondence to the problem of implementing 2-bit integer multiplication, which we will call $f : \Sigma^4 \longrightarrow \Sigma^4$, with 4 hidden spins. For simplicity we will encode the inputs and outputs as binary numbers in the obvious way, treating each spin as a bit, though it should be mentioned that more optimal number representations do exist \cite{novelrealnumberrepresentations}.  This function is large enough to be nontrivial in the sense that it is not solvable without using hidden spins. However, it is small enough that the partition function, and therefore the exact values of the spin expectations, can be computed easily, unlike, for instance, an implementation of MNIST.

We will train a neural network $\phi_{W,b}$ with $\tanh$ activation, four input neurons, four hidden neurons in one layer, and four output neurons, so that $\phi_{W,b}$ approximates $f$ with $L^\infty$ loss less than $\epsilon$. Then, we will create the Ising system corresponding to this neural network, which has free parameters $\delta$ and $\beta$. We will see, as predicted, that lowering $\delta$ with $\beta = 1$ fixed will produce a correct circuit. However, it will also be interesting to look at the behavior of the circuit as the values of $\delta$ and $\beta$ vary.

For any inverse temperature $\beta$, a choice of $\delta$ small enough will make the assumption of \Cref{eq:ansatz-directional} essentially accurate (see \Cref{lem:backwards}). However, increasing $\beta$ towards $\infty$ with a fixed small $\delta$ will not replicate $\phi_{W,b}$, but instead $\mathcal{P}_{w,b}$, the neural network with weights $W,b$ and the activation function sign (see \Cref{sec:deep}). Therefore, if $\text{sgn}(\phi_{W,b}) = \mathcal{P}_{W,b}$, then we should expect there to exist $\delta_*$ and $\beta_*$ such that the Ising circuit implements $f$ in the high temperature sense when $\beta > \beta_*$ and $\delta < \delta_*$. On the other hand, if $\text{sgn}(\phi_{W,b}) \neq \mathcal{P}_{W,b}$, we should \textit{a priori} expect the Ising circuit to implement $f$ only for $\beta$ in an interval around $1$. In the following we will present two choices of $W,b$ for this circuit which exhibit these two cases. 

\subsection{Correctness Heatmaps}

To visualize when correctness arises or does not, we need to plot the `correctness' of the circuit for varying values of $\delta$ and $\beta$. This entails a choice of a numerical measure of correctness. We may restate the high temperature notion of solution as follows: for a given input $\sigma^{(0)}$, and a choice of output bit $\sigma^{(L)}_i$, we call the bit `correct' if $\text{sgn}(\langle \sigma^{(L)}_i | \sigma^{(0)}\rangle) = f_i(\sigma^{(0)})$. It therefore makes sense to use $c_i(\sigma^{(0)}) :=\langle \sigma^{(L)}_i | \sigma^{(0)}\rangle  f_i(\sigma^{(0)})$ as a continuous measure of correctness, where $1$ indicates perfectly correct, and $-1$ indicates perfectly incorrect. A given Ising system, understood as attempting to implement a function, must get every output bit right for every input. Therefore the overall circuit performance is most effectively summarized as the \textit{worst case correctness} value $\eta$, defined as 
\begin{align}
    \eta :=\min_{\sigma^{(0)} \in \Sigma^{M_0}} \min_{i \in [M_L]} f_i(\sigma^{(0)}) \langle \sigma^{(L)}_i | \sigma^{(0)}\rangle.
\end{align}
It is easy to see that $\eta \in (-1, 1)$ and that $\eta > 0$ if and only if the circuit is correct.

The best way to visually summarize the effects of the parameters $\beta$ and $\delta$ is through a 2-dimensional heatmap of $\eta$ where the axes are the two parameters. We will use dark blue to represent a value of $+1$ and dark red to represent $-1$, with a smooth gradient between the two colors. White will represent a value of $0$, which is also of course useless, and should be considered pure noise. Blue (positive) regions will correspond to parameters which give correct circuits, and red (negative) regions to parameters which give a circuit which has at least one wrong bit for at least one input configuration. The focus on the minimum value is dictated by the application: a circuit is not much good if it reliably gets the wrong answer to even one combination of inputs; in fact, we would rather say that such a circuit is simply computing a different function than the one intended. 

\begin{figure}[h!]
    \centering
    \begin{subfigure}[t]{.45\textwidth}
        \centering
        \includegraphics[height=3in]{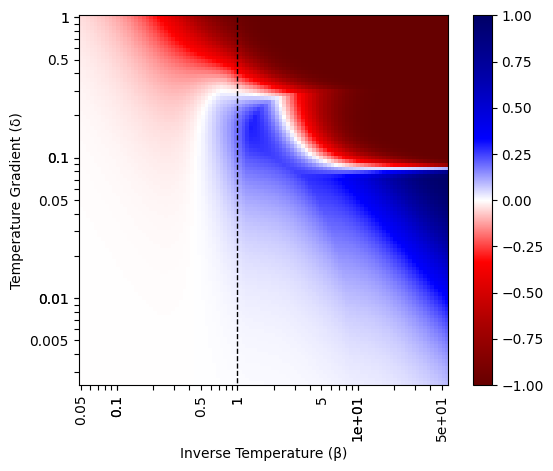}
        \caption{Model A.}
        \label{fig:phaseA}
    \end{subfigure}%
    ~
    \begin{subfigure}[t]{0.54\textwidth}
        \centering
        \includegraphics[height=3in]{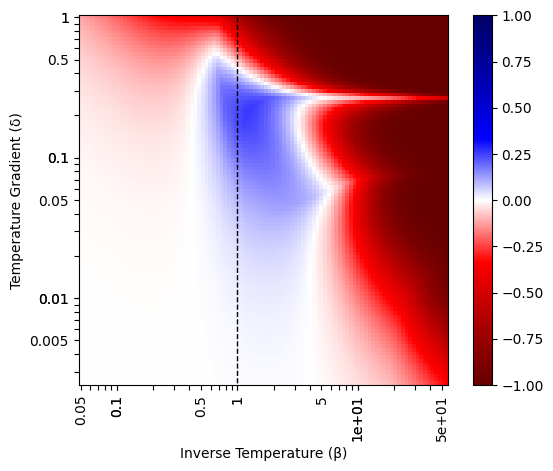}
        \caption{Model B}
        \label{fig:phaseB}
    \end{subfigure}
    \caption{Heatmap plots of worst case correctness ($\eta$) for two models implementing $2$ bit integer multiplication. Dark blue indicates $+1$ (perfectly correct) while dark red indicates $-1$ (totally incorrect). Any blue area corresponds to a correct circuit. Bright blue and bright red correspond to $\pm0.25$, since the qualitatively important differences in spin averages are concentrated around 0.}
    \label{fig:phase}
\end{figure}

We will consider two models, model A and model B. They were each trained to model the 2 bit integer multiplication function with $\epsilon = 0.05$, but with different weight initializations. In \Cref{fig:phase} the behavior of the two models is compared at varying values of $\delta$ and $\beta$. The dashed line indicates the $\beta=1$ isotherm, along which \Cref{thm:main} guarantees that the plot will eventually be blue as $\delta \rightarrow 0$. Model A has the property that $\mathcal{P}_{W,b} = f$, but for model B the opposite is true, i.e. there exists an input $\sigma$ and an index $i$ such that $\mathcal{P}_{W,b}(\sigma)_i = -f_i(\sigma)$. This difference explains the difference in qualitative behavior between the two models in the bottom right corner: model A is also a solution in the low temperature sense for $\delta < 0.08$, but model B is not a solution in the low temperature sense for any choice of $\delta$. This is visible in the plots in the differing extents of the blue (correct) region: the blue region extends infinitely far right toward the bottom left corner of the diagram for model A, but is confined horizontally by red regions in model B. In either case it is evident, as expected, that very high temperatures, whether globally or only on the output layer, result in useless noise, as indicated by the white regions. The interesting and complex structure of the plotted functions reflects the fact that $\eta$ is the minimum over a 64 individual correctness functions, each of which looks like a simple curve oriented and located differently. 

\subsection{Behavior of Individual Spins}

While $\eta$ effectively summarizes overall circuit performance, it is not entirely enlightening regarding the spin-by-spin behavior of the circuit. Therefore, it will be useful to visualize the individual correctness values $c_i(\sigma)$ for each choice of $\sigma$ and $i$ to see how the behavior of individual output spins affects overall circuit performance. In fact, it is usually only a small subset of the cases that become problematic. For a given value of $\beta$ and $\delta$ we will plot the values of $c_i$ for each $\sigma$ in a heatmap. The set $\Sigma^N$ of all inputs comes with a natural ordering given by interpreting each input as a 4-bit binary number, allowing us to denote the set of possible inputs as $x_i$ for $1 \leq i \leq 2^N$. We will make a heatmap where the square at row $i$ and column $j$ is the color corresponding to the value $c_j(x_i)$, where again dark blue indicates fully correct and dark red indicates totally incorrect. In these plots, the $i$th row shows the correctness of the output $Y$ for the input $X = x_i$. 

\begin{figure}[h!]
    \centering
    \begin{subfigure}[t]{.8\textwidth}
        \centering
        \includegraphics[width=.9\linewidth]{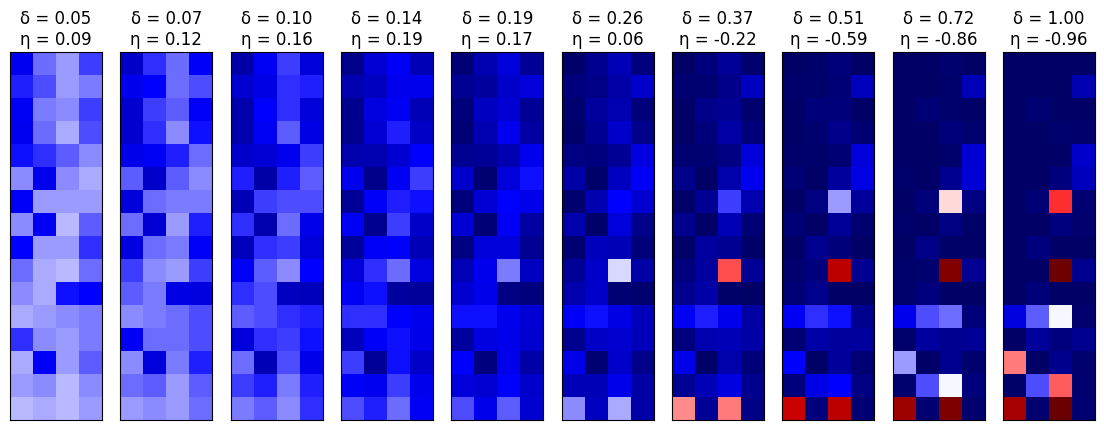}
        \centering
        \label{fig:vertA}
    \end{subfigure}%
    ~\\
    \begin{subfigure}[t]{0.8\textwidth}
        \centering
        \includegraphics[width=.9\linewidth]{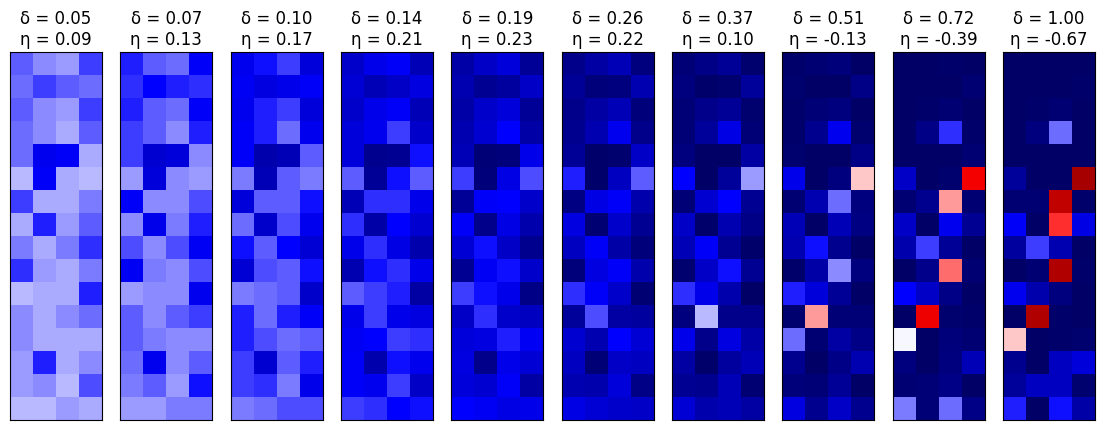}
        \label{fig:vertB}
    \end{subfigure}
    \caption{Heatmap plots of spin-wise correctness for each input pattern, plotted at several values of $\delta$ with $\beta=1$ fixed. Model A is above, model B below. The two models behave similarly, exhibiting errors with insufficient temperature gradients (right) and high noise at excessive temperature gradients (left), with an optimal zone of correctness in the center.}
    \label{fig:vert}
\end{figure}

First, we will compare the behavior of the two models along the $\beta=1$ isotherm, which is the regime covered by our main result. One can think of this as plotting a detailed slice along the dashed black line in \Cref{fig:phase}, where each heatmap is the expansion of a single pixel in the the $\eta$ plot, which is its minimum value. The results are shown in \Cref{fig:vert}, with model A on top and model B on the bottom. We observe that the best results, at $\eta=0.19$ for model A and $\eta=0.23$ for model B, are achieved at $\delta = 0.14$ and $0.19$ respectively. When $\delta$ is too small, on the left side of the figure, the plot grows white, reflecting the high degree of noise in the system. While correct, these circuits are more difficult to extract useful data from. When $\delta$ is large, as on the right side of the figure, we see backwash from the output layer contaminating the system and overwhelming the forward flow of information. Qualitatively, model B performs modestly better, but the behavior patterns of the two models are quite similar. 

\begin{figure}[h!]
    \centering
    \begin{subfigure}[t]{.8\textwidth}
        \centering
        \includegraphics[width=.9\linewidth]{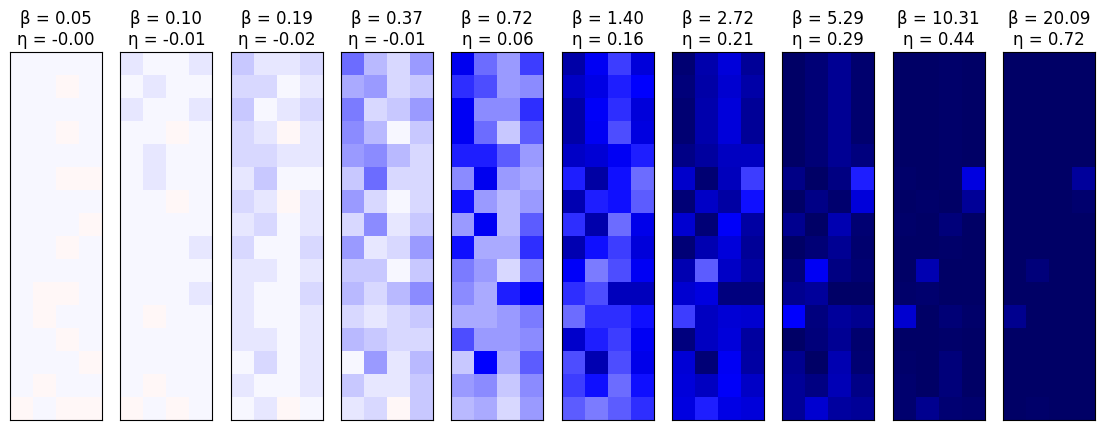}
        \centering
        \label{fig:horA}
    \end{subfigure}%
    ~\\
    \begin{subfigure}[t]{0.8\textwidth}
        \centering
        \includegraphics[width=.9\linewidth]{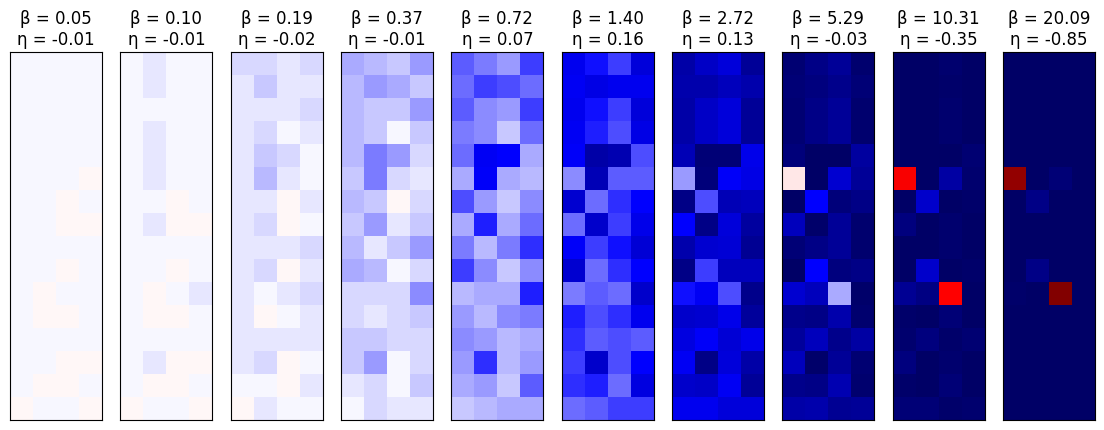}
        \label{fig:horB}
    \end{subfigure}
    \caption{Heatmap plots of spin-wise correctness for each input pattern, plotted at several values of $\beta$ with $\delta=0.07$ fixed. Model A is above, model B below. While both models are correct at intermediate temperatures (center) and noisy at high temperatures (left) at low temperatures (right) model A remains correct while model B develops errors.}
    \label{fig:hor}
\end{figure}

Secondly, we will look at the behavior of each model for several values of $\beta$ with $\delta = 0.07$ fixed. This corresponds to tracing a horizontal slice through \Cref{fig:phase}. Based on that figure, we can predict that as $\beta$ increases, model A will converge toward perfect correctness, while model B will develop errors. The results are plotted in \Cref{fig:hor}. As predicted, the difference in performance as $\beta$ increases is quite striking. High temperatures (left) result in incorrect and noisy circuits. In the `good' region around $\beta=1$, both circuits are correct. As $\beta$ increases, model A converges on a deterministic (zero-temperature) correct circuit, while model B behaves incorrectly in two cases, corresponding to the cases where $\mathcal{P} \neq f$. 

\Cref{fig:hor} conclusively shows that there exist Hamiltonians that are solutions in the high-temperature sense but not the zero-temperature sense (for a truly rigorous proof, we must apply the results of \Cref{sec:deep} to the present example). This concludes our toy example and demonstrates the practical utility of using high temperatures to increase circuit performance under the high-temperature notion of solution: in the case of model B, more noise and randomness actually led to better results (when properly controlled). This is a situation which is at the heart of the dream of probabilistic computation, and which is facially quite unintuitive, since it contradicts the natural intuition that increased entropy should always decrease information processing performance. 

\section{Shallow $\tanh$ Perceptrons are High Temperature Ising Circuits}
\label{sec:main}

The argument given in \Cref{sec:heuristic} was only a non-rigorous sketch and skipped over many subtleties. This section proves rigorously the informal result for the case of shallow networks derived in \Cref{sec:heuristic} and experimentally exhibited in \Cref{sec:shallow}.

\begin{theorem}
\label{thm:main}
    Consider the shallow $\tanh$-activated feed forward neural network $\phi_{W,b}$ with layer sizes $M_0, M_1, M_2$. Suppose that $f : \Sigma^{M_0} \lra \Sigma^{M_2}$ is a fixed Boolean function and that we have a trained network with weights $W$ and biases $b$ which models $f$ to near-perfection, i.e. for some fixed $\epsilon < 1$,
    \begin{align}
        \max_{\sigma^{(0)} \in \Sigma^{M_0}} \left\|f\left(\sigma^{(0)}\right) - \phi_{W,b}\left(\sigma^{(0)}\right)\right\|_\infty < \epsilon.
    \end{align}
    Then, there exists $\Delta > 0$ such that for any $0 < \delta < \Delta$, the Ising circuit with the same network topology as the neural network, $X = \sigma^{(0)}$, $Y = \sigma^{(2)}$, and parameters $h^{(1)} = b^{(1)}$, $J^{(1)} = W^{(1)}$, $h^{(2)} = \delta b^{(2)}$, $J^{(2)} = \delta W^{(2)}$, $\beta = 1$ satisfies $\text{sgn}(\langle Y | X=x\rangle) = f(x)$ for any $x \in \Sigma^{M_0}$, where $\langle \cdot \rangle$ indicates the average with respect to the Boltzmann distribution. In other words, this Ising system implements $f$ in the high-temperature sense.
\end{theorem}

The strategy will, as with the heuristic argument, appeal to an approximation of the Ising system by naive mean field theory, a good fit since we are working with a high effective temperature and need the hyperbolic tangent function to appear. It will be convenient to introduce the notation $m^{(\ell)}_i$ for the approximation of $\langle \sigma^{(\ell)}_i \rangle$ by naive mean field theory. To prove \Cref{thm:main}, we need to control the two approximations made in the heuristic argument. This means that we need to control both the difference between the neural network activations and the mean field solutions $m$ and the difference between the mean field solutions and the true spin averages. We will see that both of these errors are $o(\delta)$. In comparison, we will show that the magnitude of the spin-averages of the outputs is at least $\mathcal{O}(\delta)$. Since the error is asymptotically dominated by the magnitude, we obtain that $\text{sgn}(\langle \sigma^{(2)}\rangle) = f(\sigma^{(0)})$ for small enough $\delta$, i.e. that the circuit is correct. This argument is demonstrated rigorously in \Cref{sec:proof}. Note that all expectations $\langle \cdot \rangle$ and mean field estimates $m$ are implicitly conditional on a choice of $\sigma^{(0)}$. This is fine since we can prove the theorem independently on each choice of $\sigma^{(0)}$ and minimize over $\Delta$ at the very end. 

\subsection{Technical Lemmas}

Since all the lemmas will work with the naive mean-field solution, it is very useful first to establish a sufficient condition for the uniqueness of this solution. Below a critical temperature, the mean-field solution is no longer unique. It is vital to demonstrate that working with small $\delta$ will avoid these issues. This type of argument is standard in the statistical mechanics literature, but will be included for completeness. 

\begin{definition}[Naive Mean Field System for Adjusted Shallow Networks]
    Consider an input state $\sigma^{(0)} \in \Sigma^{M_0}$, a multiplier $0 < \delta \leq 1$, and an inverse temperature $\beta > 0$. Then the naive mean field system of nonlinear equations is 
    \begin{align}
        m^{(1)}_i &= \tanh\left(\beta b^{(1)}_i + \beta \sum_{j=1}^{M_0} W^{(1)}_{ij} \sigma^{(0)}_j + \beta \delta \sum_{j=1}^{M_2} W^{(2)}_{ji} m^{(2)}_j \right) && 1 \leq i \leq M_1,\\
        m^{(2)}_i &= \tanh\left(\beta \delta b^{(2)}_i + \beta \delta \sum_{j=1}^{M_1} W^{(2)}_{ij} m^{(1)}_j \right) && 1 \leq i \leq M_2.
    \end{align}
    It will be more convenient to package the information with the following definitions of the effective local field strength vector $r$ and interaction matrix $J$:
    \begin{align}
        r(s) &:= \begin{bmatrix} b^{(1)}_i + W^{(1)} s \\ \delta b^{(2)} \end{bmatrix},\\
        J &:= \begin{bmatrix}
            0 & \delta W^{(2)}\\
            \delta (W^{(2)})^T & 0
        \end{bmatrix}        .
    \end{align}
    This simplifies the system of equations to $m = \tanh(\beta (r(\sigma^{(0)}) + Jm))$, where the $\tanh$ is understood to apply to a vector element-wise. 
\end{definition}

\begin{lemma}[Uniqueness of Mean-Field Solution for Adjusted Shallow Networks]
\label{lem:unique}
    Work with the shallow network $L =2$. There exists $\beta_* > 0$ such that for any $0 < \delta \leq 1$ and $0 \leq \beta < \beta_*$, and for any $\sigma^{(0)} \in \Sigma^{M_0}$, there is a unique solution $m$ to the system $m = \tanh(\beta (r(\sigma^{(0)}) + Jm))$. The solution is also stable. 
\begin{proof}
    Define $R(m) := \tanh(\beta (r(\sigma^{(0)}) + Jm))$. Then  we may calculate the Jacobian elements as 
    \begin{align}
        (DR)_{ij} = \frac{\partial R_i}{\partial m_j} = \frac{\partial}{\partial m_j} \left[\tanh\left(\beta r_i(\sigma^{(0)}) + \beta\sum_{j} J_{ij}m_j\right)\right]\\
        = \beta J_{ij} \sech^2\left(\beta r_i(\sigma^{(0)}) + \beta\sum_{j} J_{ij}m_j\right).
    \end{align}
    Therefore, since the range of $\sech$ is $(0,1]$, we have $|(DR)_{ij}| \leq \beta |J_{ij}|$. 
    Now observe that we have a bound on the spectral radius of the Jacobian of $R$ given by the Frobenius norm:
    \begin{align}
        \rho(DR) \leq \|DR\|_F \leq \|\beta J\|_F = \beta \delta \sqrt{2} \|W^{(2)}\|_F \leq \beta \sqrt{2} \|W^{(2)}\|_F.
    \end{align}
    Then setting $\beta_* := (\sqrt{2}\|W^{(2)}\|_F)^{-1}$
    gives us $\beta < \beta_* \implies \rho(DR) \leq \beta/\beta_* < 1 \implies \sup \rho(DR) < 1$. Therefore for any $x, y \in \R^{(M_1+M_2)}$, where $\|\cdot\|$ is the Euclidean norm,
    \begin{align}
            \|R(x) - R(y)\| = \left\|\int_0^1 \frac{d}{dt} R(y + t(x-y))\, dt\right\| = \left\|\int_0^1 (DR|_{y+t(x-y)})(x-y)\, dt\right\|\\
            \leq \int_0^1 \|(DR|_{y+t(x-y)})(x-y)\|\, dt \leq \left(\sup \rho(DR) \right) \|x-y\| \int_0^1 dt < \|x-y\|.
    \end{align}
    Thus $R$ is a contraction mapping and therefore $m = R(m)$ has a unique solution by the Banach fixed point theorem; this solution is also stable in the sense of function iteration. Note that we could improve this result by multiplying $\beta_*$ by a factor of $\delta^{-1}$, but this is not necessary for the following arguments.
\end{proof}
\end{lemma}

Knowing that the mean-field result is unique for high enough temperatures, we next need an estimate of how accurately it predicts the exact spin averages $\langle s\rangle$. We tackle this through a version of the classic Plefka-Georges-Yedida expansion of a Helmholtz free energy around $\beta = 0$ \cite{plefka1982convergenceconditiontap, georges1991expandmeanfieldtheory, yedidiaidiosyncratic, Nguyen_2017}. In the traditional approach to the system, we would expand the Helmholtz free energy in $\beta$ for a fixed $h$ and $J$, then show that the error $m - \langle s \rangle$ scaled like $\mathcal{O}(\beta^2)$. However, in our setup we have two scaling factors, $\delta$ and $\beta$, so this approach will not give a useful error bound. Fortunately, \Cref{thm:main} only applies to the $\beta = 1$ case, so we can reformulate our problem as an expansion of the Helmholtz free energy in $\delta$. Note that here we assume that all biases are independent of $\delta$, while in \Cref{thm:main} we allow the biases on the output layer to scale with $\delta$. This discrepancy will be resolved in \Cref{lem:full-accuracy}.

\begin{lemma}[Accuracy of Mean-Field Approximation]
\label{lem:accuracy}
Consider an Ising system with local fields $h$ and interaction strengths $\delta W$, where $W$ is a symmetric interaction matrix, and $\delta$ is a perturbation parameter, while $\beta=1$ is fixed. Then $\|\langle s \rangle - m\|_\infty = o(\delta)$.
\begin{proof}
We begin by writing the Gibbs free energy
\begin{align}
    G(\delta,h) = \log \sum_s \exp\left(\langle h, s\rangle +\frac{\delta}{2}\langle s, W s\rangle \right),
\end{align}
from which it follows that
\begin{align}
    \frac{\partial^2 G}{\partial h_j \partial h_i} = \frac{\partial}{\partial h_j} \langle s_i\rangle = \langle s_i s_j\rangle - \langle s_i\rangle \langle s_j\rangle.
\end{align}
Therefore the Hessian of $G$ is a covariance $\Lambda$. Now, consider any vector $c$, and observe that if $\text{Var}[c^T\sigma] = 0$, then $c^T\sigma $ is the same for any $\sigma$, since every state $\sigma$ has nonzero probability at positive temperature. It follows that $c_i= 0$; therefore, if $c \neq 0$, $\text{Var}[c^T \sigma] > 0$. Note that   $\text{Var}[c^T\sigma] > 0 \iff c^T\Lambda c > \mathbb{E}[c^T\sigma]^2$. It follows that $\Lambda$ is strictly positive definite, so $G$ is strictly convex in $h$. We obtain the Helmholtz free energy, by partial Legendre transform over the local magnetic field strengths:
\begin{align}
    A(\delta,a) :=  \sup_{h'} \{\langle a, {h'}\rangle  - G(\delta, h')\} = \langle \lambda, a\rangle - G(\delta, \lambda) 
    \\
    = -\left(\log\exp(-\langle \lambda,a\rangle)+\log \ \sum_s\exp\left(\langle\lambda, s \rangle+\frac{\delta}{2}\langle s, Ws\rangle \right) \right)
    \\
    \label{eq:gibbs}
    = -\log\sum_s\exp\left( \langle \lambda, s-a\rangle+\frac{\delta}{2}\langle s, Ws\rangle \right)     .
\end{align}
Where $\lambda(\delta,a)$ is implicitly defined by $\langle s \rangle_\lambda =a$, i.e. $a-\nabla_h G(\delta,\lambda(\delta,a))=0$.
Fortunately, \Cref{eq:gibbs} is precisely the Helmholtz energy from \cite{georges1991expandmeanfieldtheory}. This allows us to skip straight to the Taylor series expansion:
\begin{align}
\label{eq:expansion}
     A(\delta,a) = \sum_i \left[\frac{1+a_i}{2}\log \frac{1+a_i}{2} + \frac{1-a_i}{2} \log \frac{1-a_i}{2}\right] - \frac{\delta}{2} \langle a, Wa\rangle + o(\delta).
\end{align}
Since $G$ is strictly convex and essentially smooth in $h$, it is Legendre type, so since $G$ is smooth, $A$ is smooth.\footnote{For $f$ of Legendre type, note that $f^*(z) = \langle z, (\nabla f)^{-1} (z)\rangle - f((\nabla f)^{-1}(z))$. If $f$ is smooth, then $\nabla f$ is smooth bijection, so $(\nabla f)^{-1}$ is also smooth. Since $f^*$ is a composition of smooth functions, it is also smooth.} Since $A$ is smooth, we may exchange the second partial derivatives, allowing us to obtain the Taylor series of $\nabla_a A$ in $\delta$ by applying $\nabla_a$ to \Cref{eq:expansion}:
\begin{align}
    h^{(a)}_i = \frac{\partial}{\partial a_i} [ A(\delta, a)] = \arctanh a_i - \delta (Wa)_i + o(\delta) \implies a = \tanh(h^{(a)} + \delta Wa + o(\delta)),
\end{align}
where a vector quantity $o(\delta)$ refers to a finite length vector where each term is $o(\delta)$, and $h^{(a)}$ denotes the inferred local field strengths from the fixed magnetizations $a$. If we return to our true system, with $h$ and $W$ fixed and $\delta$ variable, setting $a = \langle s \rangle$ must recover $h^{(a)}  = h$, so we get the equation
\begin{align}
    \langle s \rangle = \tanh(h + \delta W\langle s \rangle + o(\delta)).
\end{align}
Finally, comparing against the mean field equations for $m$ (which can be obtained by truncating the $o(\delta)$ factor) gives us the $L^\infty$ error estimate: 
\begin{align}
    \|m - \langle s\rangle\|_\infty = \|\tanh(h + \delta Wm)) - \tanh(h + \delta W\langle s\rangle + o(\delta)))\|_\infty \\
     \leq \|(h + \delta Wm) - (h + \delta W\langle s\rangle) + o(\delta)\|_\infty \\
     \leq \delta \|W(m-\langle s \rangle)\|_\infty + o(\delta)  
     \leq \delta \|W\|_\infty \|m-\langle s \rangle\|_\infty + o(\delta) \\
     \implies \|m - \langle s \rangle\|_\infty \leq \frac{o(\delta)}{1 - \delta \|W\|_\infty} = \left(\sum_{n=0}^\infty (\delta \|W\|_\infty)^n\right) o(\delta) = o(\delta)\\
     \implies \|m-\langle s \rangle\|_\infty = o(\delta).
\end{align}
Since we are working asymptotically in $\delta \rightarrow 0$, we may use $\delta < 1/\|W\|_\infty$ in order to appeal to the geometric series formula and assume that the mean field solution $m$ is unique by \Cref{lem:unique}. Both here and elsewhere in this paper, $\|\cdot\|_\infty$, when applied to a matrix, refers to the operator norm induced by the $L^\infty$ vector norm. 
\end{proof}
\end{lemma}

There is a weakness to the above error estimate: It assumes that the biases remain fixed as the interaction strengths scale. For the purposes of our model, we will need to scale some of the biases as well. However, this can be accomplished at the cost of negligible error:

\begin{lemma}[Accuracy for $\delta$-Scaled Biases]
\label{lem:full-accuracy}
    Consider an Ising system with local fields $h + \delta h_s$ and interaction strengths $\delta W$, where $W$ is a symmetric interaction matrix, and $\delta$ is a perturbation parameter, while $\beta=1$ is fixed. Then $\|\langle s \rangle - m\|_\infty = o(\delta)$.
\begin{proof}
We will convert the scaled biases to interactions with a new spin $\eta$ which has such a strong bias that it essentially always 1. Define (using block notation)
\begin{align}
    W' := \begin{bmatrix}
        W & Sh_s\\
        Sh_s^T & 0
    \end{bmatrix},
    &&
    h' := \begin{bmatrix}
        h \\ K
    \end{bmatrix},
\end{align}
where $K >> 1$ and $S := 1/\tanh(K)$ are constants.

We will write the expected values of the spins of the augmented system as $(\langle s \rangle', \langle \eta \rangle')$, where $\langle \cdot \rangle'$ denotes expectation with respect to the augmented system $(W', h')$. We would first like to show that $\|\langle s \rangle - \langle s\rangle'\|_\infty = o(\delta)$. We will introduce the partition functions $Z'$ for the augmented system and $Z$ for the original system, with the extra spin $\eta$ included but non-interacting:
\begin{align}
    Z' :=\sum_s\sum_{\eta = \pm 1}\exp\left(\langle h, s\rangle + \frac{\delta}{2} \langle s, Ws\rangle + \eta(K +  \delta S \langle h_s, s\rangle)\right),\\
    Z:= \sum_s\sum_{\eta = \pm 1}\exp\left(\langle h, s\rangle + \frac{\delta}{2} \langle s, Ws\rangle + \eta K +  \delta  \langle h_s, s\rangle\right).
\end{align}
We can calculate their difference as
\begin{align}
 Z' - Z = \sum_s\sum_{\eta = \pm 1}\exp\left(\langle h, s\rangle + \frac{\delta}{2} \langle s, Ws\rangle + K\eta\right)\left(\exp(\eta  \delta S \langle h_s, s\rangle) - \exp\left(\delta  \langle h_s, s\rangle\right)\right)
 \\
 = \sum_s\sum_{\eta = \pm 1}\exp\left(\langle h, s\rangle + K\eta\right)\left(1+\frac{\delta}{2} \langle s, Ws\rangle + o(\delta)\right)\left(\delta \langle h_s, s\rangle(\eta S - 1) + o(\delta)\right)
  \\
   = \delta \sum_s \langle h_s, s\rangle \exp \left(\langle h, s\rangle\right)  \sum_{\eta = \pm 1} e^{\eta K}(\eta S - 1) + o(\delta).
\end{align}
Now note that
\begin{align}
  \sum_{\eta = \pm 1} e^{\eta K}(\eta S - 1) = e^{K}(S - 1) + e^{-K}(-S - 1) = S\sinh K - \cosh K = 0.
\end{align}
Therefore, $Z' - Z = o(\delta)$. It follows that
\begin{align}
    \frac{Z'}{Z} = \frac{Z + o(\delta)}{Z} = 1+ o(\delta) \implies \log Z' - \log Z = \log(1+o(\delta)) = o(\delta) \implies \|\langle s \rangle' - \langle s \rangle\|_\infty = o(\delta).
\end{align}
Where we have used the fact that $Z > 0$ when $\delta = 0$, and therefore $o(\delta)/Z = o(\delta)$. 

The second step is to estimate the mean-field error. Denote the mean field solution for the augmented system with bias $h'$ as $(m', m'_\eta)$ and the mean field solution of the original system with bias $h+\delta h_s$ as $m$. The mean field equations for the augmented system read
\begin{align}
    m' = \tanh(h + \delta(Wm' + m_\eta S h_s)),\\
    m_\eta = \tanh(K + \delta\langle Sh_s, m'\rangle).
\end{align}
By the same logic as previous arguments,
\begin{align}
    \|m-m'\|_\infty \leq \|h + \delta h_s + \delta Wm - (h+\delta(Wm' + m_\eta Sh_s))\|_\infty\\
    = \delta\|(1-Sm_\eta)h_s + \delta W(m-m')\|_\infty \leq \delta \|h_s\|_\infty |1-Sm_\eta| + \delta \|W\|_\infty \|m-m'\|_\infty \\
    \label{eq:onedelta}
    \implies \|m-m'\|_\infty \leq \frac{\delta \|h_s\|_\infty}{1 - \delta\|W\|_\infty} |1-Sm_\eta|= |1-Sm_\eta|\mathcal{O}(\delta).
\end{align}
Now, using a series expansion for $\tanh$ grants
\begin{align}
    |1-Sm_\eta| \leq  1-S\tanh(K-S\delta\|h_s\|_1) = 1 - S(\tanh K - \delta S\|h_s\|_1 \sech^2 K) + o(\delta).
\end{align}
Plug in the definition of $S$ to obtain
\begin{align}
    |1-Sm_\eta| \leq \delta \|h_s\|_1 \frac{\sech^2 K}{\tanh^2 K} + o(\delta) = \mathcal{O}(\delta) + o(\delta).
\end{align}
Combining this with \Cref{eq:onedelta} gives us $\|m-m'\|_\infty \leq \mathcal{O}(\delta^2) + o(\delta) = o(\delta)$

Finally, applying \Cref{lem:accuracy} to the augmented system gives us $\|\langle s \rangle' - m'\|_\infty = o(\delta)$. Putting our estimates together, 
\begin{align}
    \|\langle s \rangle - m\|_\infty \leq \|\langle s\rangle ' - \langle s \rangle\|_\infty + \|\langle s \rangle' - m'\|_\infty  + \|m - m'\|_\infty = o(\delta).
\end{align}
\end{proof}
\end{lemma}

The final lemma concerns the accuracy of the unidirectional information flow ansatz \Cref{eq:ansatz-directional}, which for our purposes only needs a very loose bound. By scaling of the weights of the next layer, as we propose to do in the theorem, we can make the mean field solution of a given layer arbitrarily close to the corresponding $\tanh$-activated neural network layer. In \Cref{thm:main} we are concerned only with a shallow network, so this lemma will allow us to compare the hidden layer's mean field average state to the activated state of the hidden layer neurons in the feed-forward model that we are using as a training surrogate. 

\begin{lemma}[Control of Backwards Information Flow]
\label{lem:backwards}
Consider the general setup of \Cref{sec:heuristic}. Define $A_i(m) := (J^{(\ell)}m^{(\ell-1)})_i + h^{(\ell)}_i$ and $B_i(m) = ((J^{(\ell+1)})^Tm^{(\ell+1)})_i$. Assume that we have a (not necessarily unique) mean field solution $m$, which in particular gives $m^{(\ell)} = \tanh(\beta(A + B))$.
Then we can control the accuracy of the comparison to the corresponding neural network layer as
$\|m^{(\ell)} -  \tanh(\beta A)\|_\infty \leq \beta \|J^{(\ell+1)}\|_\infty$. 

\begin{proof}
First, since $m^{(\ell+1)}$ is a solution to the mean field equations, it is contained in the $\|\cdot\|_\infty$-ball of radius 1 about the origin, that is $[-1,1]^{M_{\ell+1}}$. Therefore the backward influence is bounded: 
    $\max_i |B_i| \leq \|(J^{(\ell+1)})^T\|_\infty$. 
Since $\tanh$ is 1-Lipschitz, $\|m^{(\ell)} -  \tanh(\beta A)\|_\infty
    \leq \left\|\beta B\right\|_\infty \leq \beta\|(J^{(\ell+1)})^T\|_\infty$. 
\end{proof}
\end{lemma}

\subsection{Proof of \Cref{thm:main}}
\label{sec:proof}

\begin{proof}
Fix a choice of $\sigma^{(0)}$. Suppose by \Cref{lem:unique} $\delta$ is small enough that $m$ is unique, and $\delta < 1/\|W^{(2)}\|_\infty$. We can summarize the results of \Cref{lem:full-accuracy} and \Cref{lem:backwards} as 
\begin{align}
    \|m^{(1)}-\tanh(W^{(1)}\sigma^{(0)} + h^{(1)})\|_\infty \leq \delta\|W^{(2)}\|_\infty,
    \\
    \|\langle \sigma^{(2)}\rangle - m^{(2)}\|_\infty = o(\delta).
\end{align}
By definition, $m^{(2)} = \tanh(\delta(h^{(2)}+W^{(2)}m^{(1)}))$. Substitution then gives us
\begin{align}
    \|\tanh(\delta(h^{(2)}+W^{(2)}\tanh(W^{(1)}\sigma^{(0)} + h^{(1)}))) - m^{(2)}\|_\infty
    \\
    = \|\tanh(\delta(h^{(2)}+W^{(2)}\tanh(W^{(1)}\sigma^{(0)} + h^{(1)}))) - \tanh(\delta(h^{(2)}+W^{(2)}m^{(1)}))\|_\infty
    \\
    \leq \|\delta(h^{(2)}+W^{(2)}\tanh(W^{(1)}\sigma^{(0)} + h^{(1)})) - \delta(h^{(2)}+W^{(2)}m^{(1)})\|_\infty
    \\
    = \delta\|W^{(2)}(\tanh(W^{(1)}\sigma^{(0)} + h^{(1)}) -  m^{(1)})\|_\infty
    \\
    \leq \delta\|W^{(2)}\|_\infty \|m^{(1)}-\tanh(W^{(1)}\sigma^{(0)} + h^{(1)})\|_\infty \leq \delta^2\|W^{(2)}\|_\infty^2 = o(\delta).
\end{align}
Now by the triangle inequality, we split the error into our two asymptotically bounded terms:
\begin{align}
    \|\tanh(\delta(h^{(2)}+W^{(2)}\tanh(W^{(1)}\sigma^{(0)} + h^{(1)}))) - \langle \sigma^{(2)}\rangle\|_\infty 
    \\
    \leq\|\tanh(\delta(h^{(2)}+W^{(2)}\tanh(W^{(1)}\sigma^{(0)} + h^{(1)}))) - m^{(2)}\|_\infty + \|\langle \sigma^{(2)}\rangle - m^{(2)}\|_\infty = o(\delta).
\end{align}
By definition of little-o, there exists $\Delta > 0$ such that for all $0 < \delta < \Delta$, 
\begin{align}
    \|\tanh(\delta(h^{(2)}+W^{(2)}\tanh(W^{(1)}\sigma^{(0)} + h^{(1)}))) - \langle \sigma^{(2)}\rangle\|_\infty < (1-\epsilon)\delta.
\end{align}
Now, observe the following lower bound on the magnitude, which follows from the fact that the network was trained to $\epsilon$-near perfection:
\begin{align}
    \min|\tanh(\delta(h^{(2)}+W^{(2)}\tanh(W^{(1)}\sigma^{(0)} + h^{(1)})))| \\
    \geq \delta \min |\tanh(h^{(2)}+W^{(2)}\tanh(W^{(1)}\sigma^{(0)} + h^{(1)}))| > (1-\epsilon)\delta.
\end{align}
It follows that the error never exceeds the magnitude:
\begin{align}
    \max |\tanh(\delta(h^{(2)}+W^{(2)}\tanh(W^{(1)}\sigma^{(0)} + h^{(1)}))) - \langle \sigma^{(2)}\rangle| \\
    < \min|\tanh(\delta(h^{(2)}+W^{(2)}\tanh(W^{(1)}\sigma^{(0)} + h^{(1)})))|.
\end{align}
Therefore, sign errors never occur, i.e.
\begin{align}
    \text{sgn}(\langle \sigma^{(2)}\rangle) = \text{sgn}(\tanh(\delta(h^{(2)}+W^{(2)}\tanh(W^{(1)}\sigma^{(0)} + h^{(1)}))))\\
    = \text{sgn}(\tanh(h^{(2)}+W^{(2)}\tanh(W^{(1)}\sigma^{(0)} + h^{(1)}))) = f(\sigma^{(0)}).
\end{align}
Minimizing $\Delta$ over all choices of $\sigma^{(0)}$ gives a $\Delta_* > 0$ because there are finitely many choices of $\sigma^{(0)}$. This ensures that $\delta < \Delta_*$ implies $\text{sgn}(\langle \sigma^{(2)} | \sigma^{(0)}\rangle) = f(\sigma^{(0)})$ for all $\sigma^{(0)}$, proving the theorem. 
\end{proof}

\section{Deep Binary Perceptrons are Zero-Temperature Ising Circuits}
\label{sec:deep}

The study of neural networks originates with the binary perceptron machine of the 1950s. In modern terminology, this was a shallow feed forward neural network with Heaviside activation. This is of course equivalent to using sign as an activation function, up to a linear change of variables. In this section, we consider the problem of implementing a deep binary (a.k.a. sign-activation) perceptron. This sort of model is not very common in training because backpropagation is not effective on this activation function. However, it can arise as a feed-forward network with a sigmoid or tanh activation whose activation functions have been aggressively quantized to 1 bit of precision. 

We saw in \Cref{sec:heuristic} that implementing a neural network as an Ising system with $J^{(\ell)} = \delta^{\ell-1}W^{(\ell)}$ distorts the activation functions by $\tanh(A) \to \tanh(C_\ell(\beta A+\mathcal{O}(\delta)))$, where $A$ is the incoming pre-activation to a spin from the previous layer and $C_\ell$ is a constant dependent on the layer and $\delta$. It therefore stands to reason that for small enough $\delta$, as $\beta \rightarrow \infty$ we realize the activation function $\text{sign}(A)$. Therefore, we hypothesize that deep feed forward neural networks with sign activation can be realized as zero-temperature Ising systems using our correspondence principle. This behavior was shown experimentally for a single model in \Cref{fig:hor}. In fact, it is universal. This is our second main result:

\begin{theorem}
\label{thm:deep}
Suppose $\mathcal{P}_{W,b}  = f$ for a $L$-layer feed forward neural network $\mathcal{P}$ with sign activation. Furthermore, assume that $\mathcal{P}_{W,b}$ is well defined, i.e. no activation is zero for any choice of input (since the sign function is not well defined on the input zero). Then, there exists $\delta_* > 0$ such that for any $\delta \leq \delta_*$, the Ising system with the same network topology and $J^{(\ell)} = \delta^{\ell-1}W^{(\ell)}$, $h^{(\ell)} = \delta^{\ell-1}b^{(\ell)}$ implements $f$ in the zero-temperature sense. It follows that for a choice of $\delta$ there exists $\beta_* > 0$ such that for any $\beta > \beta_*$, the system implements $f$ in the high-temperature sense. 
\begin{proof}
Fix input pattern $\sigma^{(0)} \in \Sigma^{M_0}$. Define $\sigma_*^{(k)} = \text{sgn}(W^{(k)}\sigma_*^{(k-1)}+b^{(k)})$ for all $k\leq L$ and $\sigma_*^{(0)} = \sigma^{(0)}$. Then $\sigma_*$ are precisely the activations of the deep sign network when it is fed input $\sigma^{(0)}$. All we really have to do is prove that $\sigma_*$ is the global minimum of the conditional Hamiltonian $H(\cdot|\sigma^{(0)})$. From \Cref{def:cor} we have the $\delta$-adjusted Hamiltonian
\begin{align}
    H_\delta(\sigma) = -\sum_{k=1}^L \delta^{k-1}\langle \sigma^{(k)}, b^{(k)} + W^{(k)}\sigma^{(k-1)}\rangle.
\end{align}
We will write $H$ for $H_\delta$ since the $\delta$-dependence is understood. Define $K_0 := \max_{k,v}\|b^{(k)}+W^{(k)}v\|_1$ and $K_1 =\min_{\sigma^{(0)},k,j} |(b^{(k)} + W^{(k)}\sigma_*^{(k-1)})_j|$. Note that $K_1 > 0$ by our assumption that the perceptron is well-defined. For some state $\sigma$ of the entire system, we use the notation $\sigma^{<\ell}$ to mean the vector of layer states $(\sigma^{(0)},\dots,\sigma^{(\ell-1)})$, and $\sigma^{>\ell}$ similarly. 

Now, we proceed by induction.
Suppose $H(\sigma_*^{<\ell}, \sigma^{\geq \ell}) < H(u)$ for any $\sigma^{\geq \ell}$ and $u$ such that $u^{<\ell} \neq \sigma_*^{<\ell}$. Now take any $u^{(\ell)} \neq \sigma_*^{(\ell)}$ and fix any $\sigma^{>\ell}$, $u^{>\ell}$. Additionally, for convenience use $\sigma^{(\ell)}:=\sigma_*^{(\ell)}$. Observe that
\begin{align}
    H(\sigma_*^{<\ell}, u^{(\ell)}, u^{>\ell})-H(\sigma_*^{<\ell}, \sigma_*^{\ell},\sigma^{>\ell}) 
    = 
    \delta^{\ell-1}\langle \sigma_*^{(\ell)} - u^{(\ell)}, b^{(k)} + W^{(k)}\sigma_*^{(k-1)}\rangle 
    \\
    -
    \sum_{k=\ell+1}^L \delta^{k-1}(\langle u^{(k)}, b^{(k)} + W^{(k)}u^{(k-1)}\rangle - \langle \sigma^{(k)}, b^{(k)} + W^{(k)}\sigma^{(k-1)}\rangle).
\end{align}
We can lower bound the first term above zero, showing that it is positive:
\begin{align}
    \delta^{\ell-1}\langle \sigma_*^{(\ell)} - u^{(\ell)}, b^{(k)} + W^{(k)}\sigma_*^{(k-1)}\rangle = 2\delta^{\ell-1} \sum_{j | (\sigma_*^{(\ell)})_j \neq u^{(\ell)}_j} |(b^{(k)} + W^{(k)}\sigma_*^{(k-1)})_j| \geq 2\delta^{\ell-1}K_1 > 0.
\end{align}
We can also upper bound the magnitude of the second term:
\begin{align}
    \left|\sum_{k=\ell+1}^L \delta^{k-1}(\langle u^{(k)}, b^{(k)} + W^{(k)}u^{(k-1)}\rangle - \langle \sigma^{(k)}, b^{(k)} + W^{(k)}\sigma^{(k-1)}\rangle)\right|\\
    \leq \sum_{k=\ell+1}^{L}2\delta^{k-1}\max_v\|b^{(k)}+W^{(k)}v\|_1\leq 2\max_{k,v}\|b^{(k)}+W^{(k)}v\|_1 \sum_{k=\ell+1}^L\delta^{k-1} <\frac{2K_0\delta^{\ell}}{1-\delta}.
\end{align}
Notice that 
\begin{align}
    2\delta^{\ell-1}K_1 \geq \frac{2K_0\delta^{\ell}}{1-\delta} \iff \delta \leq \frac{K_1}{K_0+K_1}.
\end{align}
Therefore, setting 
$$\delta_*=\frac{K_1}{K_0+K_1}$$ 
gives us
$H(\sigma_*^{<\ell}, u^{(\ell)}, u^{>\ell})>H(\sigma_*^{<\ell}, \sigma_*^{\ell},\sigma^{>\ell})$. 

Now, pick any $\sigma^{>\ell}$ and $u$ such that $u^{\leq\ell} \neq \sigma_*^{\leq\ell}$. If $u^{<\ell}\neq \sigma_*^{<\ell}$, $H(\sigma_*^{\leq\ell}, \sigma^{> \ell}) < H(u)$ by the inductive hypothesis. Otherwise, we must have $u^{(\ell)} \neq \sigma_*^{(\ell)}$, so by the above argument, $H(\sigma_*^{\leq\ell}, \sigma^{> \ell}) < H(u)$. Either way, the inductive step is proven. It follows that $\delta \leq \delta_*$ implies that $H(\sigma_*) < H(u)$ for all $ u \neq \sigma_*$ such that $ u^{(0)} = \sigma_*^{(0)}$, for all $\sigma^{(0)}$. It follows that for $\delta < \delta_*$, $H_{\delta}$ implements $f$ in the zero-temperature sense. Therefore, there exists $\beta_* > 0$ such that for all $\beta > \beta_*$, $(H_\delta, \beta)$ implements $f$ in the high-temperature sense. 
\end{proof}
\end{theorem}

\section{Practical Concerns: Quantization and Sparsification}
\label{sec:quantization}

Real Ising-type computing devices typically require the local magnetic fields and coupling strengths to be small integers. Some implementations, such as the coherent Ising machine (CIM) even require the couplings to be in ${-1, 0, 1}$, a special case sometimes called a `unit Ising machine'. Heretofore we have discussed Ising machines whose parameters are real numbers. Therefore, in order to make the presented design method practical for real devices, we must at least briefly address how quantization, that is the rounding of interactions to integers without affecting qualitative performance, may be integrated into our framework. Rather than starting from scratch, we can borrow existing algorithms for quantizing neural networks. Sparsification can also be achieved with existing algorithms, and may be useful for fitting Ising circuits into sparser graph topologies, but it is trivial---our methods do not affect sparsity, so starting with a sparse neural network $\phi_{W,b}$ will ensure sparsity of the result. We will therefore focus on quantization, which is more subtle. 

Getting the quantization and temperature gradient to work together correctly is not totally straightforward. To be precise, by a \textit{successful quantization of} $\phi_{W,b}$ we mean a $\tanh$-activated neural network $\phi_{Q,p}$ such that $\|\phi_{W,b} - \tilde{\phi}_{Q,p}\|_\infty < 1-\epsilon$ and the elements of $Q$ and $p$ are in $\gamma \Z$ for some $\gamma >0$. The latter condition justifies the description of `quantization' and the former ensures that $\text{sign} \circ \phi_{W,b} = \text{sign} \circ \phi_{Q,p}$, justifying the term `successful'. Having a Hamiltonian $H$ with coefficients in $\gamma \Z$ is equivalent to having an integer coefficient Hamiltonian, since we can define $\tilde{H} = \gamma^{-1} H$ and $\tilde{\beta} = \gamma$, and the Boltzmann distribution on $(\tilde{H}, \tilde{\beta})$ is the same as the distribution on $(H,\beta=1)$. 

The question now becomes whether we should quantize before or after adding the temperature gradient $\delta$. Clearly, if we quantize before multiplying the output weights by $\delta$, as prescribed in \Cref{thm:main}, we are only allowed to use $\delta = 1/k$ where $k \in \N$: notice that if $H$ has coefficients in $\gamma \Z$, $H_\delta$ has coefficients in $\gamma\delta \Z$. If we transform to integer coefficients, obtaining the system $(\gamma^{-1}\delta^{-1}H_\delta,\beta=\gamma\delta)$, we find that we have effectively multiplied the weights of the \textit{first} layer by $k$. This explains why $k$ must be integer to preserve quantization when adding the gradient, and why $k$ must be very small if we want our final interaction strengths to be small integers: if the first layer of $H$ had coefficients in $\{-\gamma, 0, \gamma\}$, the final Ising system has first layer coefficients in $\{-k,0,k\}$. It might therefore be tempting to apply $\delta$ before quantization, i.e. to quantize $H_\delta$ rather than $H$. However, this method offers no guarantee that the result will actually work, even if the chosen $\delta$ gave working Ising system before quantization. Even if the quantization is successful from the perspective of preserving the outputs of the corresponding neural networks, it can introduce unwanted backflow of information in the Ising setting. The best approach, therefore, is to apply some temperature gradient before quantization, and some temperature gradient after quantization.

\begin{figure}[h!]
    \centering
    \includegraphics[width=.9\linewidth]{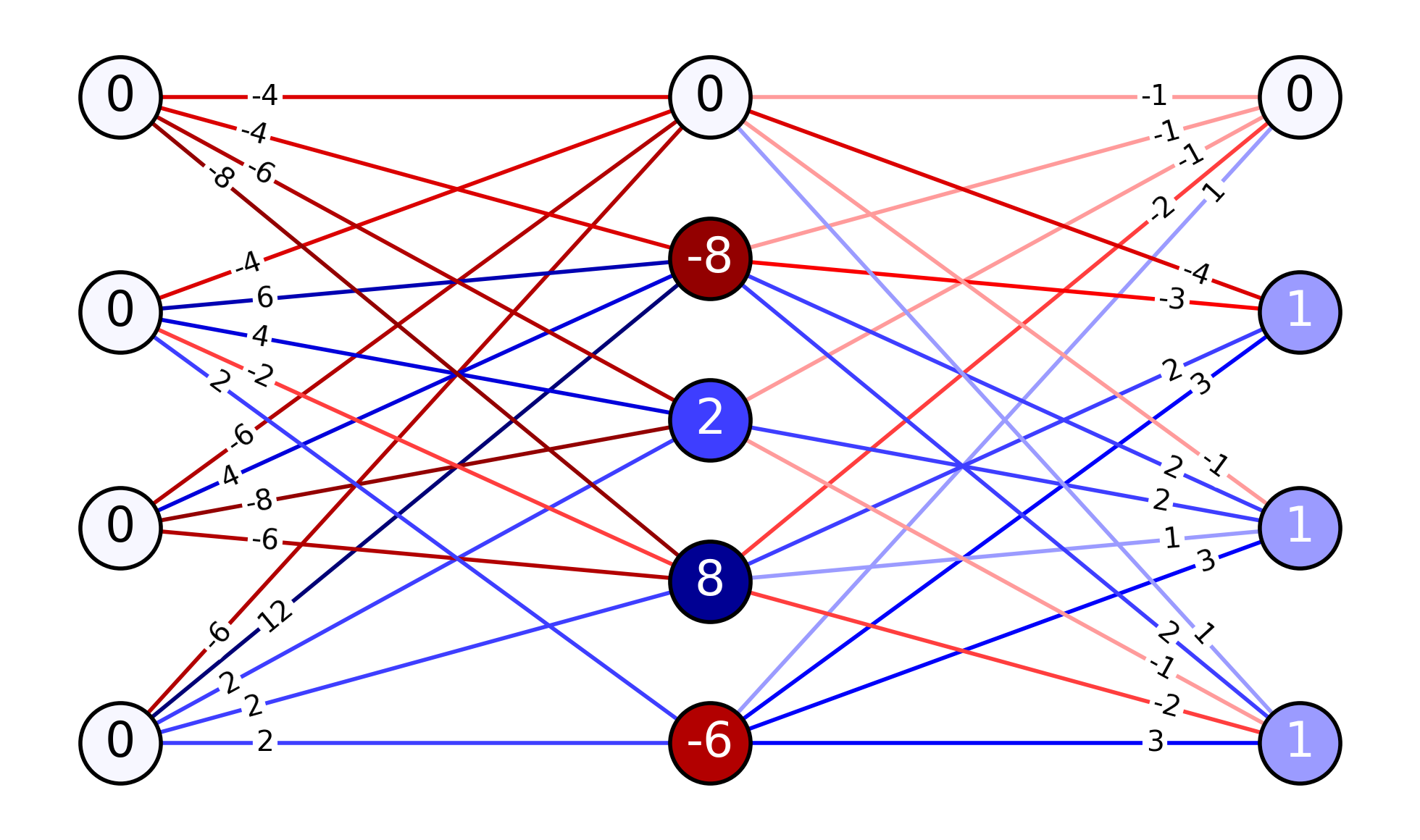}
    \caption{Quantized $2$ bit integer multiplication with integer interaction strengths. Inputs are the left column, outputs right. Node labels are the biases on each spin.}
    \label{fig:quant}
\end{figure}

We will consider as an example a $2$-bit integer multiplication circuit with $5$ hidden spins, comparable to our experiments from \Cref{sec:shallow}. The final result is plotted in \Cref{fig:quant}. We applied QRONOS quantization, an improved version of OPTQ with strong theoretical guarantees \cite{zhang2025provableposttrainingquantizationtheoretical}, to the quantization of $\phi_{W,b}$ with an applied temperature gradient of $0.5$. After quantization with $\gamma=0.5$, we applied another temperature gradient of $0.5$, then scaled the system to integers. The final result is a working multiplication circuit with small integer interaction strengths and biases, lying well within the range of $[-16, +16]$ required by many currently existing experimental devices. The final circuit operates at $\beta=0.25$ and has a minimum correctness value of $\eta=0.16$, indicating a working circuit without excessive noise. In comparison, the non-quantized Ising circuit corresponding to $\phi_{W,b}$ with $\delta=0.25$ achieves $\eta=0.22$ at $\beta=1$, indicating that we have retained about $3/4$ of the initial performance after the quantization procedure. This experiment demonstrates that effective quantization yielding results of contemporary practicality is possible with existing techniques and within the framework of \Cref{thm:main}. However, the choice of procedure was \textit{ad hoc}. It may be possible to use the recently discovered $L^\infty$ and $L^2$ bounds for neural network quantization performance, combined with analysis of Ising systems, to predict working quantization hyperparameters without compute intensive trial and error. This is left for future work. 

\section{Conclusions}

The experiments in \Cref{sec:shallow} and \Cref{sec:quantization} were conducted on small systems, both so that the microscopic behavior could be visualized easily and so that the Boltzmann expectations could be computed exactly. However, the main advantage of neural networks is that they scale well with problem size, defeating the curse of dimensionality that plagues the zero-temperature linear programming approach discussed in \Cref{sec:intro}. This implies that the true value of this paper's results is that they allow, in principle, the implementation of very large neural networks on Ising hardware, at the cost only of network training and temperature gradient selection. In the future, this could be used to run AI models on ultra-low-power edge hardware, or to execute complex inference directly in a quantum annealer. However, a major unknown remains: we do not know how well these methods scale to much larger circuits. We would like to see future work validate the performance of a large pretrained neural network mapped directly to a D-Wave quantum annealer or a classical thermodynamic sampling unit (TSU) using the methods introduced here, but there are a number of hurdles that must be jumped before our theory can be turned into practice in this way. 

First, and most obviously: how do we actually choose $\delta$? The primary drawback of \Cref{thm:main} is that it only covers the qualitative behavior of a single slice of the correctness plot in \Cref{fig:phase}. It does not construct the value of $\delta$ that optimizes $\eta$ along the $\beta=1$ isotherm. Furthermore, it is evident from the figure that the best choice of parameters $(\delta,\beta)$ is often not even on the $\beta=1$ curve, rendering it quite outside the grasp of the theoretical methods employed in our analysis. In practice, therefore, the results \Cref{thm:deep} and \Cref{thm:main} are mostly useful in that they guarantee the existence of a solution parameterized by $(\delta,\beta)$ and tell the experimenter roughly where to find it. Using a surrogate for the Boltzmann distribution, we can use nonlinear optimization to select a $(\delta,\beta)$ that approximately maximizes $\eta$, employing the theorems to intelligently select an initial guess. The relatively well-behaved shape of the $\eta$ surface plotted in the figures suggests that gradient descent may be extremely effective for this optimization. In any case, this process is much less computationally intensive than designing the circuit directly with integer-linear programming. 

Shallow neural networks, especially large ones, are surprisingly powerful, and we believe that their implementation offers a lot of value. However, the revolution in artificial intelligence is inseparable from deep learning. Ultimately, the temperature gradient method has inherent limitations in this regard. As we discussed in \Cref{sec:heuristic}, the temperature gradient method cannot faithfully represent high temperature $\tanh$-circuits with more than one hidden layer. While \Cref{thm:deep} does in theory allow for arbitrarily deep networks, this is hampered in practice by the fact that most hardware places fairly strict dynamic range restrictions on interaction strengths. In  \cite{sequentialspinlogic} deep Ising circuits with linear rather than geometric interaction strength scaling were introduced, but the methods employed there relied on the exact design of the logic gates. It remains to be seen if anything of this nature is possible for our more general circuits. In any case, the practical implementation of deep neural networks directly in Ising hardware remains an unsolved challenge. 

What about wider, instead of deeper? Advances in the theory of mean field neural networks (MFNN) and neural network Gaussian processes (NNGP) promise great mathematical tractability in the large-width limit of shallow neural networks. This brings about a natural continuation of the present research: we have applied statistical mechanics to small finite systems, but left out the thermodynamic limit characteristic of most actual physics. What about the limit of large systems, where the hidden layer size greatly increases? What do the Lee-Yang zeroes of a shallow neural network look like? Do shallow neural networks interpreted as Ising systems exhibit interesting phase transitions in the Gaussian process limit? What is the $\delta$-$\beta$ phase diagram of a trained system of infinite size? We hope in future that the answers to these questions will prove interesting and practical both in the engineering of Ising circuitry and the theory of machine learning. 

\section{Acknowledgments}

The author gratefully acknowledges support through the NSF RTG grant DMS-1840314 and NSF grant DMS-2009800, supervised by Professor Thomas Chen, University of Texas at Austin. The author would also like to thank his collaborator John Daly at the Laboratory for Physical Sciences. 

\printbibliography

\end{document}